\documentclass[12pt,preprint]{aastex}
\def\deg{\hbox{$^\circ$}}
\newcommand{\etal}{{\it et al. }}
\newcommand{\asec}{$^{\prime\prime }$}

\begin{document}
\title{Optical Spectroscopy of the Surface Population of the $\rho$ Ophiuchi Molecular Cloud: The First Wave of Star Formation}

\author{Bruce A. Wilking}
\affil{Department of Physics and Astronomy, University of Missouri-St.  Louis\\
1 University Boulevard, St.  Louis, MO 63121\\ bwilking@umsl.edu}

\author{Michael R. Meyer}
\affil{Steward Observatory, The University of Arizona, Tucson, AZ 85721 \\
mmeyer@gould.as.arizona.edu}

\author{John G. Robinson}
\affil{Department of Physics and Astronomy, University of Missouri-St.  Louis\\
1 University Boulevard, St.  Louis, MO 63121\\ johnr@newton.umsl.edu}

\and

\author{Thomas P. Greene}
\affil{NASA/Ames Research Center, M.S. 245-6 \\
Moffett Field, CA  94035-1000 \\
tgreene@mail.arc.nasa.gov}

\begin{abstract}
We present the results of optical spectroscopy of 139 stars obtained with the Hydra
multi-object spectrograph.  The objects extend over a 1.3 square degree
area surrounding the main cloud of the $\rho$ Oph complex.  The objects 
were selected from narrowband images to have H$\alpha$ in emission.
Using the presence of strong
H$\alpha$ emission, lithium absorption, location in the Hertzsprung-Russell diagram,
or previously reported x-ray emission, we were able to identify 88 objects as
young stars associated with the cloud.  Strong H$\alpha$ emission was confirmed
in 39 objects with line widths consistent with their origin in magnetospheric
accretion columns.  Two of the strongest emission-line objects are young, x-ray emitting
brown dwarf candidates with M8 spectral types.  Comparisons of the bolometric
luminosities and effective temperatures with theoretical models suggest a median
age for this population of 2.1 Myr which is signifcantly older than the ages
derived for objects in the cloud core.  It appears that these stars formed
contemporaneously with low mass stars in the Upper Scorpius subgroup, likely triggered
by massive stars in the Upper-Centaurus subgroup.  

\end{abstract}

\section{Introduction}

Nearby molecular clouds which are the sites of active star formation are observed 
to be surrounded by a population of less obscured 
and more evolved 
young stars.
H$\alpha$ emission-line and x-ray surveys have revealed large numbers of
pre-main sequence stars that extend well beyond the molecular cloud boundaries.
For example, the ROSAT All-Sky Survey has led to the identification of 112
lithium-rich pre-main sequence stars that are spread over a 450 square degree area
surrounding the Orion star-forming molecular clouds (Alcal\'a \etal 1996).  
While some of the young stars appear to be related to earlier star formation in
Gould's Belt, at least 60\% are found directly associated with the Orion molecular cloud complex
(Alcal\'a \etal 2000).  Proposed origins for these associated young stars include
formation in
rapidly moving cloudlets that have since dispersed (Feigelson 1996) or ejection
from the cloud core due to three-body encounters (Sterzik \& Durisen 1995).
Alternatively, they could simply be fossil tracers of previously exisiting dense
molecular gas in analogy with OB associations and now dispersed giant molecular clouds (e.g., Blaauw 1991).
Clearly, determining the age and distribution of pre-main sequence stars surrounding the
denser molecular gas holds the
key to describing the star formation history of a region and perhaps sets limits to
the lifetime of the molecular cloud.

The $\rho$ Oph cloud complex, located the the edge of the Upper Scorpius subgroup in
the Sco-Cen OB association, is comprised of a series of filamentary dark clouds that 
extend eastward from cores of dense molecular gas (de Geus 1992).  
At a distance of about 150 pc, it is one of the closest regions of active star formation.
The main cloud, L1688, hosts a 1 pc x 2 pc centrally condensed molecular core where the visual
extinction is estimated to be 50-100 mag (Wilking \& Lada 1983).  The core of L1688 has been the focus
of numerous near-infrared, far-infrared, and millimeter 
continuum surveys, as well as x-ray and radio continuum surveys,
and is found to host an embedded infrared cluster with around 200 stars (e.g., 
Greene \& Young 1992; Bontemps \etal 2001; 
Andr\'e \& Montemerle 1994; Gagn\'e, Skinner, \& Daniel 2004).  Infrared
spectroscopic studies infer 
a median age of 0.3 Myr for objects in the core (Greene \& Meyer 1995;
Luhman \& Rieke 1999).  It has been suggested that star formation in the $\rho$ Oph core is 
the latest in a chain of events that began with massive stars in the Upper Centaurus-Lupus
and Lower Centaurus-Crux subgroups of the Sco-Cen OB association triggering the formation of OB stars 
in the Upper Scorpius subgroup, which in turn initiated star formation in the L1688 cloud
(de Geus 1992). 

Until recently, there has been limited study of the nature of lightly extincted association 
members at the periphery of the L1688 cloud that have been 
identified through H$\alpha$ objective prism and x-ray surveys
(e.g., Dolidze \& Arakelyan 1959; Wilking, Schwartz, \& Blackwell 1987; Montmerle \etal 1983; Casanova \etal 1995). 
Early spectroscopic studies of H$\alpha$ emission-line stars presented spectral types for about
20 young stars (Struve \& Rubkj\"{o}bing 1949; Cohen \& Kuhi 1979; Rydgren 1980).
An optical spectroscopic survey of $\rho$ Oph x-ray
sources was conducted by Bouvier \& Appenzeller (1992) and established spectral
types for 30 pre-main sequence stars (23 new).  They found an age of 1-10 Myr for these objects, 
significantly older than
the age of objects in the core (see also Greene \& Meyer 1995).    
But these samples favored the brighter association members and gave the incorrect impression
that there was a deficiency of M spectral-type stars relative to other nearby star-forming
regions (see Fig. 24, Hillenbrand 1997).  This was remedied in part by Mart\'{\i}n \etal (1998) who obtained
optical spectroscopy for 59 ROSAT-selected sources in the core and streamer of the
cloud with the majority of objects classified as pre-main sequence M stars.  By examining the ratio of 
classical T Tauri stars (CTTS) to weak-emission T Tauri stars (WTTS), they 
found evidence for an older pre-main sequence population outside of the L1688 core.

In an effort to identify more widely-distributed and/or older late-type association members,
we have obtained deep narrowband images
of a one square degree area encompassing the L1688 cloud and
centered at the wavelengths of H$\alpha$ and [S~II].  Shorter exposures at
R and I were also obtained.  
This resulted in a sample of 282 candidate CTTS with R$<$20 mag that had possible H$\alpha$ emission.
This sample includes association members with greater obscuration and/or lower mass than
targeted by previous surveys. 
Using the multi-object spectrograph Hydra,  
optical spectra were obtained for 129 candidate CTTS plus 10 additional objects with R$<$19.1 mag.
Our sample selection procedure and the observations are described in Sec. 2. 
In Sec. 3, we detail our criteria for spectral classification and our examination of
surface gravities.  The results of the spectral classifications
and an analysis
of the emission-line spectra is given in Sec. 4.  Also in Sec. 4, we list the criteria used
to identify 88 association members from the sample and their locations in a Hertzsprung-Russell diagram.
In Sec. 5, we compare the properties of the
association members with those of low mass stars in the Upper Scorpius subgroup of the
Sco-Cen OB association to gain insight into the star
formation history of the $\rho$ Oph cloud.

\section{Observations and Data Reduction}

Moderate resolution spectra were obtained for 129 of 282 stars identified
as candidate CTTS from narrowband H$\alpha$ images of the $\rho$ Ophiuchi 
cloud or from H$\alpha$ objective prism plates (Wilking \etal 1987).
In addition, spectra were obtained for 5 bright association members and 5
brown dwarf candidates in the cloud not identified as candidate CTTS.  
A total of 139 stars were observed.
These observations are described in detail in the sections below. 

\subsection{Sample Selection}

The majority of candidate CTTS were selected from deep optical images of a one 
square degree area 
centered on the L1688 cloud obtained with the Curtis Schmidt 
Telescope at Cerro Tololo Interamerican Observatory.
The images were obtained on 1995 March 10 and their reduction are described elsewhere 
(Wilking et al. 1997).  
Briefly, long (45 minute) exposure
narrowband images centered
at the wavelengths of H$\alpha$ (FWHM=64 \AA) and [S~II] (FWHM=45 \AA) were obtained in addition to 
short (5 minute) exposures in the standard R and I bands. 
The latter were calibrated in the Johnson-Kron-Cousins photometric system 
using standard star
fields established by Landolt (1992). 
R and/or I band photometry were obtained for over 
2700 stars in the Ophiuchus field with R$<$20.  Accurate positions ($<$0.5\asec)
for these stars were obtained using the ASTROM program distributed by the 
Starlink Project and a set of secondary astrometric standards.  The secondary position
references were 27 H$\alpha$ emission line stars with accurate positions 
determined relative to SAO stars in a 5 square degree region on the Red Palomar 
Sky Survey plate (Wilking \etal 1987).  
The narrowband images were utilized to estimate the
equivalent width of possible H$\alpha$ emission.  The continuum
in the H$\alpha$ filter was estimated using the flux in the [S~II] filter after correcting for the 
differences in filter widths and transmission and 
assuming a flat continuum with no [S~II] 
emission.  The latter assumption was shown to be valid, with detections of weak [S~II] emission
in only two sources.  As a result, 282 stars with R$<$20 were identified with possible 
H$\alpha$ emission [EW(H$\alpha$)$>$ 10 \AA]. 
However, because of structure in 
the continuum due to reddening and photospheric absorption bands, it was
desirable to confirm this emission spectroscopically.
   
\subsection{Hydra Observations}

One hundred fifty-one spectra were obtained for 139 stars located toward the cloud and its periphery
on 1999 May 5-6 
using Hydra, the multi-fiber 
spectrograph, on the WIYN 3.5m telescope\footnote{The WIYN Observatory is a joint facility 
of the University of
Wisconsin-Madison, Indiana University, Yale University, and the National Optical
Astronomy Observatory.}. The Bench Spectrograph Camera 
was used with the red fiber cable, the 600 lines/mm grating with a blaze angle of
13.9$\deg$ and the GG-495 
filter to cover the range of 5820-8700 \AA\ centered near 7200 \AA.  
The spectral dispersion was 1.39 \AA\ per pixel giving an effective 
resolution of 2.88 \AA. The resolution at the central wavelength was $\lambda$/$\Delta\lambda$=2500.
The program objects were selected by R magnitude, the H$\alpha$ equivalent width estimated 
from the narrowband images, and their accessibility to a fiber.
Program objects were distributed among four 
fiber configurations sorted by R magnitude; 
twelve program stars were observed in two 
configurations.  Field 1 included the brightest candidate CTTS (R$<$15) 
and, to make most efficient use of the spectrograph, 5 bright x-ray sources
known to be association members but with no
known H$\alpha$ emission.  Fields 2 and 3 sampled the same magnitude range 
of candidate CTTS (15$<$R$<$17)
with the pointing centers shifted in declination. 
Field 4 candidates (17$<$R$<$19.1) were supplemented by 5 brown dwarf 
candidates from the study 
of Wilking, Greene, \& Meyer (1999).  
We observed 81\% of the candidate CTTS
with R$<$15, 88\% of the candidates with 15$<$R$<$17, 45\% with 17$<$R$<$18,
and 21\% of the candidates
with 17$<$R$<$19.1. 
In Table 1, we summarize the observations by presenting for each field
the pointing center, R magnitude range, number of candidate CTTS
observed, integration time, and number of exposures.

The spectra were reduced using the Image Reduction and Analysis Facility
(IRAF).\footnote{IRAF is distributed by NOAO for AURA, Inc.}
Images were processed for bias and dark corrections 
using CCDPROC.  Multiple exposures of a given field were median-combined
and then reduced with IRAF's DOHYDRA package. The images 
were flat-fielded using dome flats obtained for each fiber configuration
except for Field 1 where the flat from the configuration for Field 2
was used. Sky subtraction was accomplished using the median of 7-10 sky spectra
distributed across each field.  The spectra were wavelength calibrated using 5 second exposures
of CuAr lamps.  Scattered light corrections were not made 
and no flux calibration was performed.  It was necessary in a few cases 
to improve the sky subtraction by first scaling the median sky spectrum before 
subtracting it from the source spectrum.  The typical signal-to-noise for the spectra extracted
in each field is presented in Table 1.

\section{Analysis of the Spectra}

Spectral types were derived from visual classification (visual pattern 
matching of our smoothed program star spectra
with standard star spectra) supported by 
quantitative analysis of some spectral indices.
In this section, we provide descriptions of the absorption lines used
to classifiy three broad groups, starting with the earliest spectral type stars (B-A),
moving to the F-K stars, and finally the M stars.   
We conclude the section with a discussion of gravity-sensitive absorption features in
the 5820-8700 \AA\ spectral range.
For the purposes of matching
spectral features with those of standard stars, our
Hydra spectra were
smoothed using a gaussian filter to the resolution of the standard stars for direct comparison.
All spectra have been
normalized to 1 by dividing out a fit to the continuae,
carefully excluding regions with emission lines or broad absorption due to TiO and VO.
Normalized spectra smoothed to a resolution of 5.7 \AA\ are shown in Fig. 1
for a representative sample of program objects, with early-type stars in
Fig. 1a (B3-G9), K stars in Fig. 1b, early-to-mid
M stars in Fig. 1c, and mid-to-late M stars in Fig. 1d.  Both photospheric and telluric spectral features
are labeled.  

Two main sets of standards were used for classification (both qualitative and
quantitative).  First, optical spectra from 
the WIYN/Hydra study of the Praesepe by Allen \& Strom (1995) were used 
to derive spectral types from
B8V - M4V.  The effective resolution of these spectra was 5.7 \AA.
For giants and later type dwarfs (M5V - M9V), 
optical spectra from the study of Kirkpatrick, Henry, \& McCarthy (1991) were used with an effective
resolution of either 8 or 18 \AA.  In addition to these, optical spectra of very late type subgiants
in IC~348 (Luhman 1999)
and $\rho$ Oph (Luhman, Liebert, \& Rieke 1997) were used for comparison with
the coolest stars in our sample.
For stars ealier than B8V, we referred to the spectral atlas of Torres-Dodgen and Weaver (1993).

\subsection{Classification of B-A Stars}

Absorption lines from the Balmer (n=2) and Paschen (n=3) series of hydrogen are prominent in the
spectra of early type stars.  In our spectra, we see unblended absorption lines from H$\alpha$ (6563 \AA) and 
Paschen 14 (8598 \AA) (see Fig. 1a) that reach a maximum around A0 and weaken in warmer stars.
The Ca~II triplet (8498 \AA, 8542 \AA, 8662 \AA) is also observed and decreases in strength
toward early-type stars until overtaken by Pa~16, 15, and 13 (8502\AA, 8545\AA, and 8665\AA).
Hence an F2 star
may have a similar EW(H$\alpha$) as a B5 star, but will be distinct by displaying stronger 
absorption from the Ca~II triplet.  We note that all of the aforementioned lines can appear in emission,
and that the observed absorption line strengths could be lower limits to the true strengths.
We estimate that the uncertainties in our derived spectral types in this range is $\pm$5 subclasses,
however, our adopted spectral types will be greatly aided by previously published observations
of these intrinsically bright stars.

\subsection{Classification of F-K5 Stars}
 
The relative strength of absorption due to H$\alpha$ and a blend of Ba~II, Fe~I, and Ca~I 
centered at 6497 \AA\ is the most sensitive indicator of spectral type over this range. 
Also helpful in spectral classification is
absorption due to Ca~I at 6122 \AA\ which grows stronger from A to K spectral types
(e.g., compare Figs. 1a and 1b).
Absorption from the Ca~II
triplet is present but shows little variation with spectral type in F-K stars.
In general, the variation in these features is very gradual from subclass to subclass and, when
coupled with the signal-to-noise of the spectra,
the uncertainties in the spectral classifications between F0 and K0 are often
$\pm$3-5 subclasses.
Paschen line absorption (13 and higher transitions) will be present through mid-F stars, but
Pa~13, 15, and 16 are blended and dominated by the lines from the Ca~II triplet at our resolution.
Once again,
the partial filling of H$\alpha$ absorption
by H$\alpha$ emission in CTTS could lead us to assign spectral types that are too late.

\subsection{Classification of K5-M9 Stars}

For stars K5V and later, the depths of the TiO and VO bands provide the most sensitive indicators of
spectral type. 
As shown in Fig. 1b, TiO absorption bands begin to appear in mid-K stars centered at 6300 \AA, 
6700 \AA, and 7140 \AA\ and increase in strength through M6V stars (Fig. 1c). 
Absorption due to TiO, which begins near the strong telluric O$_2$ band (``oxygen B band", 7594-7685 \AA)
and extends to 7861 \AA, is evident for spectral types of M1V and later and is further strengthened by VO
absorption from 7851-7973 \AA\ in mid-to-late M stars (Fig. 1d).  TiO absorption bands from 8432-8452 \AA\ 
also appear for spectral types
of M2V and later and increase in strength through M9V (Fig. 1d).
For the coolest M stars (M6-M9), the depths of the TiO bands centered at 7140 \AA\ and 7800 \AA\
appear to decrease, likely due to depletion of Ti into grains 
(e.g., Allen 1996, Fig. 2a); 
VO absorption from 7334-7543 \AA\ 
and from 7851-7973 \AA\ then becomes a more reliable indicator of spectral type.  

Spectral classification in this temperature range was accomplished through a combination
of visual classification and quantitative analysis of TiO and VO indices 
which we now describe.
The indices were computed by
taking the ratio of the average flux in a $\sim$50 \AA\ wide band of the continuum to the average in
an absorption band.  Following Allen (1996), we plotted the TiO index of 7035\AA/7140\AA\ vs.
the TiO index of 7500\AA/7800\AA\ along with their one sigma errors for each Hydra field 
as well as our dwarf standards.  Fig. 2a shows a plot of the TiO indices for the dwarf standards.  There
is a general trend for TiO index to increase with later spectral type until M7, where the trend reverses
and the index decreases for stars cooler than M7.  To resolve the ambiguity caused by this reversal, 
we developed two narrow band VO indices for spectral classification of the coolest stars in our sample.
A plot of these indices is shown in Fig. 2b for M dwarf standards.
The first index is the ratio of the continuum averaged over two bands which bracket
the VO band centered at 7485 \AA\ to the average flux in the VO band.  
The second index is the ratio of the continuum 
in a band centered on 8120 \AA\ to the flux in the VO band centered at 7979 \AA.
As shown in Fig. 2b, there is a general trend for the indices to increase 
from spectral types M3 to M9.
While the use of these indices allowed us to sort objects by temperature, the final spectral
classifications were made by direct comparisons of the strengths of the TiO and VO bands with
standard star spectra.  The sensitivity of these absorption bands to spectral class allowed us 
in most cases to estimate
M spectral types to within a subclass.

\subsection{Surface Gravities}

A rough estimate of the surface gravity of an object is important in determining its nature
and in estimating accurately its effective temperature for a given spectral type.
In particular, surface gravity indicators can help identify background giants and field
dwarfs that might contaminate our sample.
Gravity-sensitive absorption features
available for analysis in our spectra include 
the CaH band centered at 6975 \AA\ and the Na~I doublet at 8183/8195 \AA\
(both strongest in dwarfs) and absorption from the Ca~II triplet at 8660 \AA\ (strongest in giants).
Of these features, 
CaH is well-positioned for study.

Following Allen (1996), we have calculated a CaH index as
the ratio of the continuum at 7035$\pm$15 \AA\ to the flux in the CaH absorption band at 6975$\pm$15 \AA\
and a TiO index as the ratio of the continuum at 7030 $\pm$15 \AA\ to the flux in the TiO
absorption band at 7140$\pm$15 \AA\ from the normalized spectrum of each program object.
In Fig. 3, we plot the CaH index vs. the TiO index for 136 program objects.  Error bars are
computed based on the one-sigma error in flux averages and propagated to the ratios.
The solid lines represent first or second order fits to the standard star spectra.  Based on this plot,
we conclude that the surface gravities for objects with the largest TiO index (coolest stars)
are intermediate between
those of giants and dwarfs.  While there are clearly late-type giants present in our sample,
about 70\% of the program objects have TiO indices $<$ 2 (about spectral type M4~V) and
surface gravities that most closely resemble dwarf stars.  Therefore, 
spectral types for our program objects were determined by comparing them to a grid of dwarf standards.

\section{Results}

Spectral types were determined for 131 of 139 stars in the manner described in Sec. 3.3.
Using the CaH index and other gravity sensitive features (Sec. 3.4), six 
stars are identified as giants and 26 as possible dwarfs.
These data are presented
in Table 2 along with any previous source names, x-ray associations, RA and DEC in J2000, 
and R and I magnitudes.  
The derived spectral types (or range of possible spectral types) are also shown to the right
of each spectrum in Fig. 1 along with its classification as a possible dwarf (V?), giant (III), association member (A), 
or association unknown (U).  The top two spectra in Figs. 1c and 1d demonstrate the different features
present in high and low surface gravity objects of the same spectral type.
In the following sections, we discuss what the spectra reveal about the characteristics of the
young stars and their evolutionary states.  Included in this discussion is
the distribution of spectral 
types in our sample and the variety and properties of their emission lines.  A description is given of 
how association membership was determined and the resulting
spatial distribution of association members relative to the molecular gas.
Finally, the evolutionary state of the population is estimated 
by deriving ages and masses using a Hertzsprung-Russell diagram.

\subsection{Distribution of Spectral Types}

The distribution of spectral types for our sample is shown in Fig. 4.  
Over half (79) of the 131 objects
with spectral-type determinations are M stars.  The fact that our sample is dominated by M stars 
is due in part
to a selection effect caused by the onset of TiO absorption bands.  The procedure described
in Sec. 2.1 was designed to select objects with H$\alpha$ emission provided the continuum
was relatively flat.  But the same procedure could erroneously identify M stars with
strong TiO absorption as H$\alpha$
emission-line objects since the H$\alpha$ filter bandpass
lies in a spectral region free of TiO absorption while the comparison [S~II] filter bandpass lies
within strong TiO absorption at 6700 \AA.

The shaded areas of the histogram represent sources with strong H$\alpha$ emission
(EW(H$\alpha$)$>$ 10 \AA) at the time of our observations.  
Our procedure of selecting
H$\alpha$ emission line stars from narrowband
images was only moderately successful.  In particular, our procedure appears to break down
for the fainter sources (R$<$17) which accounts for the number of background G stars
in our spectroscopic sample.
Our highest success rate occurred with mid-K stars and mid-to-late M stars and was 65\%.  
In the latter group, 
H$\alpha$ emission was clearly a factor in their identification in our original narrowband
image.  There is clearly no deficiency of M-type CTTS in the $\rho$ Oph cloud 
relative to other nearby star-forming regions.  
The apparent lack of M stars
in previous spectroscopic surveys represents the bias toward brighter stars in those surveys.

\subsection{Emission-Line Spectra}

Nearly one-third of our sample for which we could determine spectral types displayed 
strong emission from H$\alpha$.  The spectra for about one-third of these strong emitters 
also exhibited emission lines from He~I, O~I, Ca~II, and/or Pa 14.  Emission lines were
identified and their equivalent widths and full-widths at half maximum (FWHM) were measured using
gaussian fits.  No attempt was made to deconvolve the emission profile by the instrument response.
The results for all CTTS are presented in Table 3. 

Since our sample was based on a procedure designed to select sources with 
strong H$\alpha$ emission characteristic of CTTS, it is
not surprising that 
39 of the 131 objects with spectral-type determinations displayed strong
H$\alpha$ emission.  Of these, 26 are newly identified CTTS based on an EW(H$\alpha$) $>$ 10 \AA\
criteria.  The previously identified CTTS can be found in Table 2 by looking at sources
with previously measured H$\alpha$ emission (column 10).
An additional 17 objects showed weaker H$\alpha$ emission (10 \AA\ $>$ EW(H$\alpha$) $>$ 5 \AA)
with nearly all having spectral types from M3-M5.
We note that variability plays a role in the identification of H$\alpha$ emission
(e.g., Wilking \etal 1987; Hartigan 1993). 
Among the six H$\alpha$ emitters with multiple observations, the equivalent width
of Object 2-30 (field number-aperture from Table 2) dropped
from 38 \AA\ to 8.5 \AA\ in only one day.  The brown dwarf candidate GY~5 was observed to have
an EW(H$\alpha$) = 15 \AA\ whereas a similar 
spectrum obtained in 1998 showed no detectable emission (Wilking \etal 1999).
In addition, the shape of the
emission profile also appeared to change in three cases (Chini~8, EL~24, and Object 2-30).
All of the strong H$\alpha$ emission lines
were well-resolved with typical FWHMs of 250 km s$^{-1}$ compared to
the 128 km s$^{-1}$ velocity resolution of our data.  

These broad H$\alpha$ profiles, which can exceed 300 km s$^{-1}$
in our sample, have been successfully modeled as arising in magnetospheric accretion columns
(Hartmann, Hewett, \& Calvet 1994; Muzerolle, Calvet, \& Hartmann 2001).  Thirteen of the H$\alpha$ profiles
were clearly asymmetric or displayed non-gaussian wings.  The profile shapes are noted in Table 3 using
the classification scheme of Reipurth, Pedrosa, \& Lago (1996).  The prevalence of wings or
asymmetries on the blue-side of the profile is typical among CTTS and
suggests the presence of mass outflows.  
The Ca~II triplet at 8660 \AA\ was observed in emission in 14 of the H$\alpha$ emitters
and was strongest in the G or K spectral type objects.  No attempt was made to deblend them from 
the weaker Paschen emission lines Pa~13, Pa~15, and Pa~16.  In general, the Ca~II lines were narrower
than H$\alpha$ and barely resolved in our data with typical FWHMs of 150 km s$^{-1}$.
Fig. 5 shows an expanded view of the H$\alpha$ emission and Ca~II emission lines for a representative
sample of 7 CTTS.
As in Fig. 1, the spectra have been smoothed to a resolution of 5.7 \AA.  H$\alpha$
profiles for objects 1-24 and 1-29 were symmetric while the rest of the profiles
in Fig. 5a exhibited a blueward asymmetry.
Only the source Object 2-31/1-35 (bottom two spectra in Fig. 5a and 5b)
displayed asymmetric profiles in both Ca~II and H$\alpha$.

Emission lines from both permitted and forbidden atomic transitions were detected in the spectra of the
strongest H$\alpha$ emitters in Table 3 with EW(H$\alpha$) $>$ 28 \AA.
Both Pa~14 and O~I (7773 \AA) displayed broad profiles with FWHMs similar to that of
H$\alpha$.  The other O~I lines and the He~I lines all appeared to be barely resolved, with FWHMs between
150-180 km s$^{-1}$.  The detection of forbidden line emission from [O~I] was observed
in 6 objects. Weak [S~II] emission at 6731 \AA\ was detected in only 2 objects (SR~22 and WSB~46).  
The detection of these lines
is consistent with the presence of winds associated with CTTS (Edwards \etal 1987).

\subsection{Identification of Pre-Main Sequence Association Members}

Pre-main sequence objects (PMS) objects that are products of recent star formation in the
Ophiuchus region were identified using one or more of the following four criteria.
First, candidates with H$\alpha$ emission with EW $>$ 10 \AA\ resemble CTTS, a property
exhibited by 39 stars in our study (see col. 11, Table 2).  
Second, association with x-ray emission is a signpost of youth
and has been observed in 52 stars from our sample by various surveys of the region (see
col. 4, Table 2).
Third, the presence of lithium absorption is an indicator of youth.  While the lithium absorption was 
not well-resolved in our spectra, it was definitely detected in 30 stars (col. 9, Table 2).
Finally, 58 objects (excluding giants identified in Sec. 3.3) that 
are too luminous to be main sequence objects at the distance to $\rho$ Oph 
{\it and} 
have an estimate for A$_v$ too high to be foreground to the cloud (A$_v$ $>$ 1.5 mag) are identified
as pre-main sequence objects.
Table 4 lists the 88 association members identified by our observations followed by the
criteria used in their classification.  At least 25 of these objects are
newly identified association members.  The bright association member SR~12 has been
included in list; while we did not obtain a spectrum, it was present in our R and I band 
images and had a previously well-determined spectral type.

We note that of the 26 objects with dwarf-like surface gravities (Table 2), 19 are identified as
association members based on criteria 2, 3, or 4.  None display strong H$\alpha$ emission.
Without performing a quantitative analysis, the strengths of their CaH and
Na I absorption lines are indistinguishable from main sequence stars.

\subsubsection{Lithium}
The presence of Li~I absorption at 6707 \AA\ in stellar spectra indicates the objects
have cool interiors ($<$2 x 10$^6$ K) and are young.
Measurements proved to be difficult for most objects
due to our spectral resolution and neighboring iron lines.
The line was definitely observed in 30 stars with higher S/N spectra (``yes" in col. 9, Table 2)
and tentatively detected in 19 stars (``yes:" in col. 9, Table 2).
Six stars were identified as young objects solely on the basis of lithium absorption (Table 4) and all
were located above the main sequence in the H-R diagram.
The absence of an Li entry in Table 2 or Table 4 does not imply
that Li~I is not present, only that the feature was not discernable in our
spectra.

\subsubsection{Brown Dwarfs}

This program included 7 brown dwarf candidates identified through low resolution infrared
spectroscopy:  
GY~3, GY~5, GY~37, GY~59, GY~204, GY~310, and GY~326 (Wilking, Greene, \& Meyer 1999, hereafter WGM99; 
Luhman \& Rieke 1999; Natta \etal 2002).
The new optical spectral types agreed with the infrared classifications
within 2 subclasses for 6 out of 7 sources.  
Using infrared spectra, GY~310 was classified as an M8.5 by WGM99 and Wilking, Meyer, \& Greene (2005), 
and M6 by Natta \etal (2002).
While the spectrum is noisy, our optical classification of M4 for GY~310 is significantly earlier
and, if confirmed, may suggest the presence of an infrared companion.

The two coolest objects in our sample, 4-41 (GY~3) and 4-45 (GY~264), are classified with M8 spectral types.
Their spectra, shown in Fig. 1d, feature
strong VO absorption at 7334-7534 \AA\ and 7851-7973 \AA.
The shorter wavelength VO forms a obvious depression in the continuum between two TiO
bands.
The longer wavelength VO broadens the absorption from 
the 7666-7861 TiO band starting around
7900 \AA.  Their classifications as young brown dwarf candidates was confirmed by their
strong H$\alpha$ emission [EW(H$\alpha$) =
140 \AA\ and 155 \AA, respectively], association with x-ray emission (see Table 2),
and location in the H-R diagram (Sec. 4.4).
Object 4-41, also known as ISO Oph 32, was first identified as a low mass object by Natta \etal (2002).
Excess emission at K ((r$_k$) = 0.34) and in the mid-infrared (Bontemps \etal 2001) are indicative
of a circumstellar disk (Natta \etal 2002)
\footnote{Using JHK photometry from the 2MASS catalog converted into the CIT photometric system, 
r$_k$ = F$_{Kex}$/F$_K$ = [(1+r$_h$)(10$^{[(H-K)-(H-K)_0-0.065*A_v]/2.5}$)]-1 where
r$_h$ = F$_{Hex}$/F$_H$ = (10$^{[(J-H)-(J-H)_0-0.11*A_v]/2.5}$-1 with instrinsic colors and visual
extinctions computed for an M8 spectral type.}.
Object 4-45 is a previously unidentified brown dwarf candidate.  It was previously misidentified
as a foreground M star based on VRI photometry (ophmd5, Festin 1998).  It does not appear to have
an infrared excess with r$_k$ = 0.02.  The object lies in close proximity to the ``marginal"
Herbig-Haro candidate P2 (Phelps \& Barsony 2004).

\subsubsection{Distribution of Association Members}

The distribution of the 88 association members identified by this study is shown in Fig. 6 relative to
contours of $^{13}$CO column density which delineate the cloud boundaries.  
We note that due to the high extinction in the cloud core, our
data samples only the surface population toward the densest gas.  For reference, star symbols
mark the locations of the Sco-Cen B star $\rho$ Oph~A and the the most massive member of the 
L1688 embedded cluster, Oph S1.
There is a definite tendency for assocation members to
concentrate near the highest column density gas in the core.  Only 14 members are
seen in projection outside the lowest contour corresponding to N$_{LTE}$(13) = 6$\times$10$^{14}$ cm$^{-2}$.

\subsection{Hertzsprung-Russell Diagram}
With the benefit of the spectroscopic data, we can investigate the ages and masses of the association members
identified by this study.
The intrinsic colors, bolometric corrections, and temperature scale for dwarf stars were derived
from the works of
Schmidt-Kaler (1982) for B8-K5 stars and from Bessell (1991) for K5-M7 stars.
We note that the assumption of dwarf, rather than subgiant, surface gravities 
will overestimate T$_{eff}$ for stars with spectral types of G5-K5 by $<$250 K 
and underestimate T$_{eff}$ for stars with spectral types of M2-M8 by $<$200 K.
For the B3 star Source~1, an effective temperature of T$_{eff}$ = 19,000 and an L$_*$ = 1500 L$_{\sun}$
was adopted (e.g., Lada \& Wilking 1984).  For the M8 brown dwarf candidates, we assumed values of
T$_{eff}$ = 2500, $(R-I)_0$ = 2.5, and BC(I) = -1.7 that are consistent with recent studies of very
low mass stars (Dahn \etal 2002; Hawley et al. 2002).
Bolometric luminosities were computed from the I band magnitudes in the following way.
Magnitudes were dereddened using
the (R-I) color excess derived from the observed minus instrinsic (R-I) color for 
the corresponding spectral type and using
the reddening law A$_v$/E(R-I) = 6.25 (Cohen \etal 1981).  
In a few cases the errors in the photometry and/or spectral classification yielded slightly negative
values for the extinction and an extinction of 0.0 was assumed.
Absolute I magnitudes were then computed 
assuming a distance of
150 pc (de Zeeuw \etal 1999) and converted to bolometric magnitudes using the appropriate bolometric
correction for the observed spectral type.  The stellar luminosity was computed using
Log(L/L$_{\sun}$) = 1.89 - 0.4 $\times$ M$_{bol}$ assuming M$_{bol\sun}$ = 4.74 (Livingston 1999).
Due to the dominance of emission lines in the spectrum of WL~18, we adopted the luminosity estimated
by Bontemps \etal (2001) from the J band flux and (J-H) color, adjusted to a distance of 150 pc.

The resulting H-R diagram is presented in Fig. 7 for 84 association members 
relative to the theoretical tracks and isochrones
of D'Antona \& Mazzitelli (1997, 1998).  Excluded from the diagram is the B3 star Source 1, the A7
star VSSG~14, 
and objects GY~59 and GY~326 which lack R band photometry and therefore estimates for 
A$_v$ and Log (L).
CTTS are marked by the solid diamonds; open diamonds mark association members with
weak or no H$\alpha$ emission that would correspond to weak-emission or post T Tauri stars.
The errors in T$_{eff}$ are estimated to be $\pm$105 K for K-M stars with equal contributions from the
uncertainties in the spectral classification and systematic offsets in the temperature scale.  The uncertainty in Log(L) 
is estimated to be 0.12 dex, assuming uncertainties in the R and I photometry of 0.04 mag,
in the distance modulus of 0.15 mag,
and in the bolometric correction of 0.1 mag.  The masses and ages interpolated from
the models are given in Table 4 with uncertainties due to random errors of 20\% in mass and 0.25 dex in Log(age).
Comparisons with the Baraffe \etal (1998) models suggest systematic errors between different sets
of models of $\sim$50\% in mass and $\sim$0.5 dex in Log(age) (e.g., Hillenbrand \& White 2004).

The median age of the association members in Fig. 7 is 2.1 Myr.  
There is no significant difference observed in the median age of the CTTS and the
non-CTTS association members.  Following the analysis of Preibisch \& Zinnecker (1999), the apparent age
spread observed in Fig. 7 is consistent with all of the association members having roughly the same age 
(1-3 Myr) with
the scatter in the age accounted for by the errors noted above and unresolved binaries.

\section{Implications for the Star Formation History}

The association members identified in this study are pre-main sequence objects distributed 
over a 1.3 square degree area (6.8 pc$^2$) and lie within a projected radius of 2 pc 
of the highest extinction regions of the
main $\rho$ Oph cloud (see Fig. 6).  The fact that our derived visual extinctions are 
generally low (80\% of the association members have A$_v$ $\le$ 5 mag) indicates that
the majority of our sample lies at the surface of the cloud.
Recent spectroscopic studies of YSOs in the $\rho$ Oph cloud have focussed on the 
embedded population in the 1 pc x 2 pc
centrally condensed core and have necessarily been conducted at near-infrared wavelengths 
(Greene \& Meyer 1995; Luhman \& Rieke 1999; WGM99; Natta \etal 2002).  These studies have consistently
derived ages between 0.1-1 Myr when using the D`Antona \& Mazzitelli tracks and 
isochrones, with a median age
of 0.3 Myr.  After considering our uncertainties,
we conclude that the distributed population is significantly older than that in the more 
highly extincted cloud core.

The question then arises as to the origin of this more distributed population.  One explanation 
is that these objects formed in the dense core and have diffused or been ejected
to the surface of the cloud.  However, the space motion
of a YSO which acquires the velocity dispersion of the molecular gas in which it formed 
(1-2 km s$^{-1}$) would not be high enough to move an object to the projected distances
from the core of $>$1 pc that we observe for some association members.
Alternatively, the ejection of
very low mass stars and brown dwarfs from small groups has been proposed for the origin 
of widely distributed WTTS and of substellar objects
(Sterzik \& Durisen 1995; Reipurth \& Clarke 2001; Bate \etal 2002).  In this case, one 
would predict that the lowest mass association members in our 
sample would display a higher velocity dispersion (and thus a broader spatial
distribution) than YSOs in the core. 
We have analyzed the distribution of the higher mass association members 
(M$>$0.5 M$_{\sun}$) relative the the lower mass members (M$<$0.5 M$_{\sun}$) using a
two-dimensional Kolmogorov-Smirnov (K-S) test.  The K-S statistic is consistent with 
the two populations being drawn from the same parent population.
Our data do not lend support to the ejection hypothesis, however our 
sample includes only 8 brown dwarf candidates.\footnote{Radial velocity measurements 
are required for a detailed analysis of the kinematics of these stars to reject the hypothesis that
they were ejected from the cloud as the result of the dissolution of small stellar groups.}

A second possibility is that the association members at the cloud edges are near their 
birthsites and that the dense gas in L1688 was once distributed over a larger area
than we see today.  
It has been proposed that star formation in the Upper Scorpius was triggered 
about 5 Myr ago by the passage of
an expanding HI shell driven by massive stars in the Upper Centaurus-Lupus association 
(de Geus 1992).  This age is consistent with that of high mass stars in Upper Sco 
derived from a color-magnitude diagram and from the main
sequence turnoff (Preibisch \etal 2002; see models by Bertelli \etal 1994).
The implied age of the low mass stars in Upper Sco is also 5 Myr, but only after
corrections are made for unresolved binaries 
(Preibisch \& Zinnecker
1999).  A comparsion of Fig. 7 with H-R diagrams for low mass members of
the Upper Scorpius subgroup, using the same set of theoretical evolutionary tracks
and isochrones and not corrected for unresolved binaries, shows no discernable difference 
in age or age spread (e.g., Preibisch \& Zinnecker 1999;
Preibisch, Guenther, \& Zinnecker 2001; Preibisch \etal 2002).  This is not too
surprising given that the $\rho$ Oph cloud is ringed by high-mass members of
Upper Scorpius such as the multiple star $\rho$ Oph and $\sigma$ Sco to the west and
22 Sco, $\alpha$ Sco, and $\tau$ Sco to the east. 
Therefore, one 
could regard the distributed population we observe as either low mass members of the 
Upper Scorpius subgroup or as older members of the $\rho$ Oph cloud
whose formation was contemporaneous with stars in the
Upper Scorpius subgroup.  In either scenario, ultraviolet radiation and the passage 
of an expanding shell 
from massive stars in the Upper Scorpius subgroup within the last 2.5 Myr
would have stripped away the outer skin of the $\rho$ Oph cloud, revealing the present 
population and perhaps triggering a second episode of star formation in the
in the centrally condensed core (de Geus 1992).  

\section{Summary}

We have analysed optical spectra for 139 candidate CTTS that extend over a 1.3 square degree
area in the direction of the main $\rho$ Oph cloud.  Of the 131 stars for which spectral types 
and surface gravities are estimated, 6 are identified as giants and 79
are classified as M stars.  Association with the cloud is 
established for 88 objects by the presence of optical emission lines or lithium absorption 
in the spectra, 
by location in the H-R diagram, and/or by previously reported x-ray emission.  Thirty-nine of the stars
display strong H$\alpha$ emission (EW$>$10\AA) characteristic of CTTS.  These emission lines
are broad (FWHM$\sim$250 km s$^{-1}$), consistent with their origin in magnetospheric
accretion columns.  In addition, asymmetries in the H$\alpha$ profiles and the 
presence of forbibben lines are observed in 16 objects are a signpost of 
mass outflows.  A subset of the CTTS also display emission from the Ca II triplet which is 
strongest in the most massive CTTS and, in one instance, self-absorbed.
Two of the strongest emission-line objects are young, x-ray emitting brown dwarf candidates 
with M8 spectral types.

Association members are distributed in the lower extinction regions surrounding the 
high extinction core.  A H-R diagram suggests a median age of 2.1 Myr with an apparent
age spread due mainly to uncertainties in the calculated luminosity and the presence
of unresolved binaries.  This age is significantly older than the median age
of 0.3 Myr for objects in the high extinction core.  There is no difference
in median age between the CTTS and non-CTTS association members and there are no
statistically significant correlations between the age or mass of an association member
and its projected distance from the core center.  The age and distribution of this
population is indistinguishable from low mass stars in the Upper Scorpius subgroup.  We
propose that these stars formed in what was 
once a larger $\rho$ Oph cloud. 
In this instance, star formation in the outer regions
of the $\rho$ Oph cloud would have been triggered by the same event that initiated star
formation in Upper Scorpius.

\acknowledgments

We thank Di Harmer for obtaining the Hydra data throught the WIYN Queue 
Observing Program and for assistance with the data reduction.
Lori Allen provided helpful insight into the spectral 
typing process and made available the bulk of the spectral standards
used in this study in digital form.  We gratefully acknowledge Eric Mamajek for helpful
discussions.  MRM acknowledges support for this work through a Cottrell 
Scholar's Award from the Research Corporation.
BW and JR acknowledge support from NSF AST-9820898 to the University of Missouri-St. Louis
and from the Missouri Research Board.

\pagebreak

\pagebreak

\begin{center}
{\bf Figure Captions}
\end{center}
\medskip
%
%
\figcaption{%
A collection of spectra which is representative of the entire sample is presented,
smoothed to a resolution of 5.73 \AA\ and normalized
using a polynomial fit.
The spectra are labeled with their field/aperture numbers (Table 2)
and a spectral type or range.
In addition,
indicator
of association membership (A) or undetermined (U) follows the spectral type.  Atomic and molecular bands of
interest
in our classification scheme are labled as well as telluric features.  Fig. 1a shows B-G stars, Fig. 1b shows K stars,
Fig. 1c the early M stars and Fig. 1d the late M stars.  Included in
Figs. 1c and 1d are spectra of giants and dwarfs from our survey to illustrate the intermediate surface gravity of the YSOs.
}

%
%
\figcaption{%
A plot of the spectral indices for dwarf standards used to guide our specral classifications.
Spectra from the studies of Allen \& Strom (1995) or Kirkpatrick, Henry, \& McCarthy (1991) were used in the analysis.
Fig. 2a shows a plot of the TiO index of 7035\AA/7140\AA\ vs.
the TiO index of 7500\AA/7800\AA\ for K5-M9 dwarfs.  The ratios were formed by averaging the flux in
a 30 \AA\ wide band centered on the wavelength.
Fig. 2b plots the two VO band indices which are sensitive to spectral type for late M dwarf standard stars.
Spectra from the studies of Allen \& Strom (1995) or Kirkpatrick, Henry, \& McCarthy (1991) were used in the analysis.
The width of the wavelength bands in the continuum were chosen to maximize the number of channels
yet avoid strong absorption due to TiO.
The y-axis shows the ratio between the continuum averaged between a 20 \AA\ wide band
centered at 7380 \AA\ and a 30 \AA\ wide band centered at 7555 \AA\ to the flux in a
50 \AA\ wide band centered
in the VO band at 7465 \AA.  The x-axis shows the ratio of the continuum in a 40 \AA\ wide
band centered at 8120 \AA\ to the flux in a 60 \AA\ wide band centered in the VO band
at 7970 \AA.
}
%
%
\figcaption{%
A plot of the CaH vs. TiO indices as defined in the text for 136 program objects.  The solid lines were
derived from first or second order fits to dwarf and giant spectral standards.  For the dwarf standards from K5-M7,
the fit was  y = 0.126x +0.940 with a correlation coefficient of r=0.94.
We note that for spectral types later than M7, both indices decrease in response
to an overall depression of the continuum so that an M8 V star has a CaH index similar to that of a M4 V star.
For the giant standards from K5-M5,
the fit gave y = -0.0357x$^2$ + 0.191x + 0.795 with a correlation coefficient of r=0.82.
}
%
%
\figcaption{%
The distribution of spectral types derived for 131 program objects.  The bins were chosen to
permit a direct comparison with similar histograms derived for other star-forming regions
(e.g., Hillenbrand 1997).  The shaded areas indicate objects which displayed strong
H$\alpha$ emission EW(H$\alpha$) $>$ 10 \AA.
}
%
%
\figcaption{%
An expanded view of H$\alpha$ and Ca~II emission lines in 7 CTTS labeled by their field/aperture
numbers (Table 2) and spectral types.  The bottom two spectra are from the same source observed on two different
nights.  Fig. 5a shows the structure of the
H$\alpha$ emission line with the rest wavelength of H$\alpha$ marked as a reference.  Fig. 5b shows the Ca~II
triplet for the same objects.}
%
%
\figcaption{%
The distribution of association members is shown relative to contours of $^{13}$CO column density.  The contours were
computed from Loren (1989) assuming LTE and T$_{ex}$ = 25 K.  
The values of the contours in units of cm$^{-2}$ are 6$\times$10$^{14}$,
3$\times$10$^{15}$, and 1.5$\times$10$^{16}$; the lowest contour delineates the outer boundary of the dark cloud.
The dashed box outlines the field included by our Hydra obsrvations.  Star symbols mark the locations of the star
$\rho$ Oph A (labeled) and the embedded association member Oph S1 in the L1688 core.}
%
%
\figcaption{%
A Hertzsprung-Russell diagram for the $\rho$ Oph association members with optically
determined spectral types.  The solid diamonds mark the positions of CTTS (EW(H$\alpha$ $>$ 10 \AA) and
the open diamonds correspond to WTTS or post-T Tauri stars    
relative to the theoretical tracks
of D'Antona \& Mazzitelli (1997,1998).  
Isochrones shown as solid lines are
10$^5$, 3 $\times$ 10$^5$, 10$^6$, 3 $\times$ 10$^6$, 10$^7$, and 10$^8$ years.
Evolutionary tracks
from 0.02 M$_{\sun}$ to 2.0 M$_{\sun}$ are shown by dashed lines.
The bold dashed line marks the evolutionary track for a star at the
hydrogen-burning limit.
The typical error bar
for a candidate is shown in the lower left of each plot and is
$\pm$0.015 dex in Log(T$_{eff}$) and $\pm$0.12 dex in
Log(L$_{bol}$/L$_{\sun}$).
}

\newpage
\centerline{FIGURE 1.A} 
\plotone{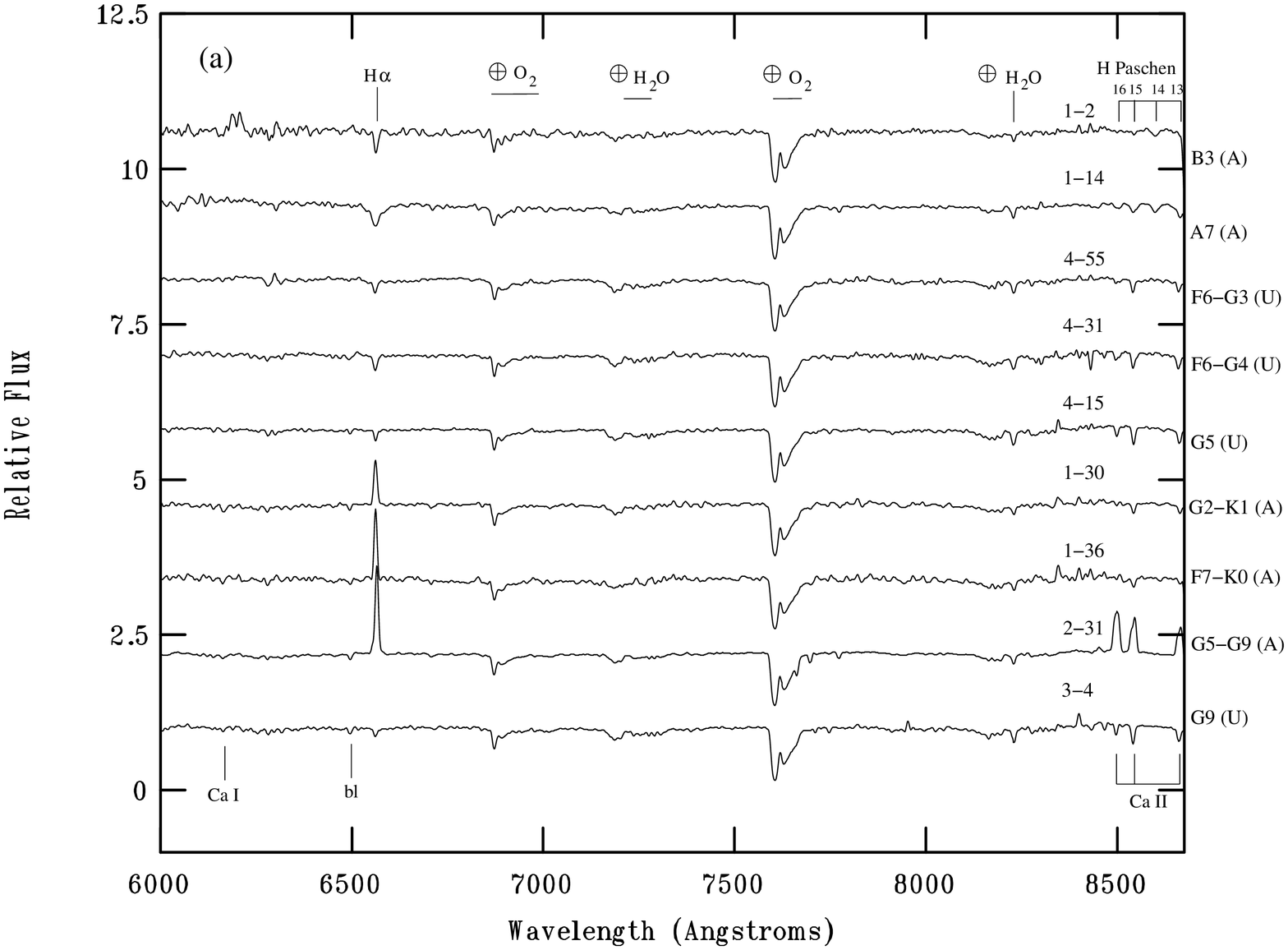}
\newpage
\centerline{FIGURE 1.B} 
\plotone{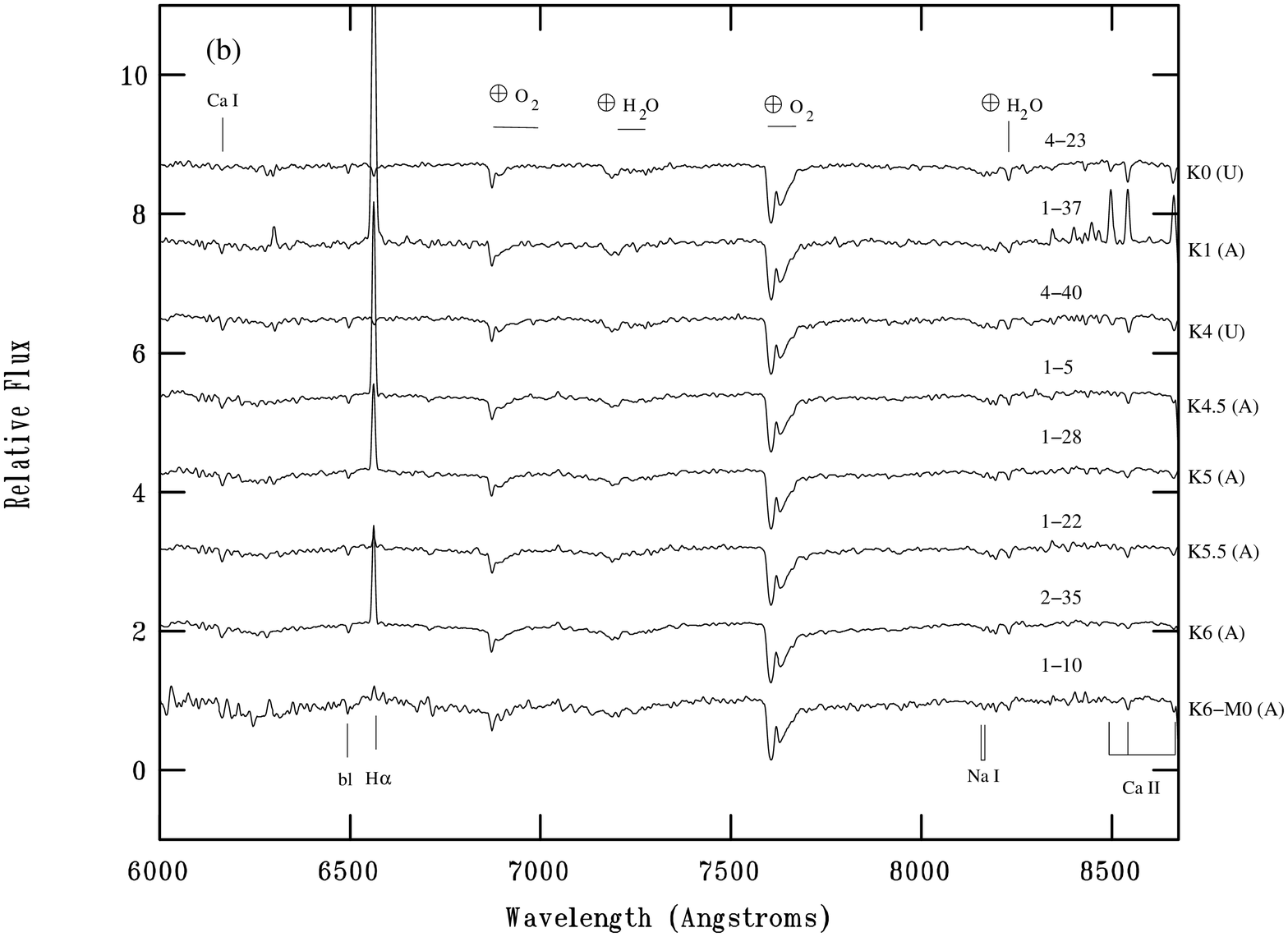}
\newpage
\centerline{FIGURE 1.C} 
\plotone{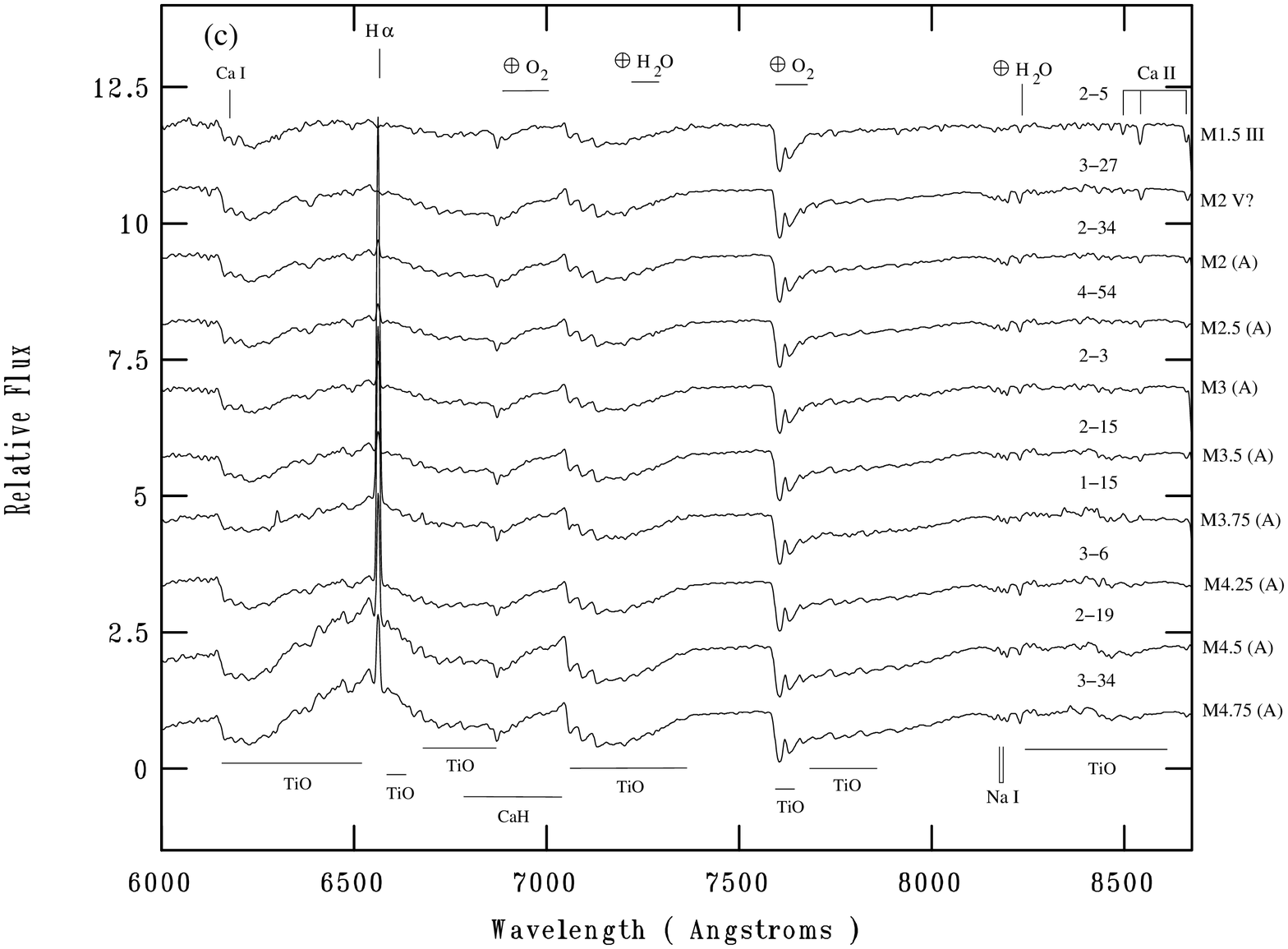}
\newpage
\centerline{FIGURE 1.D} 
\plotone{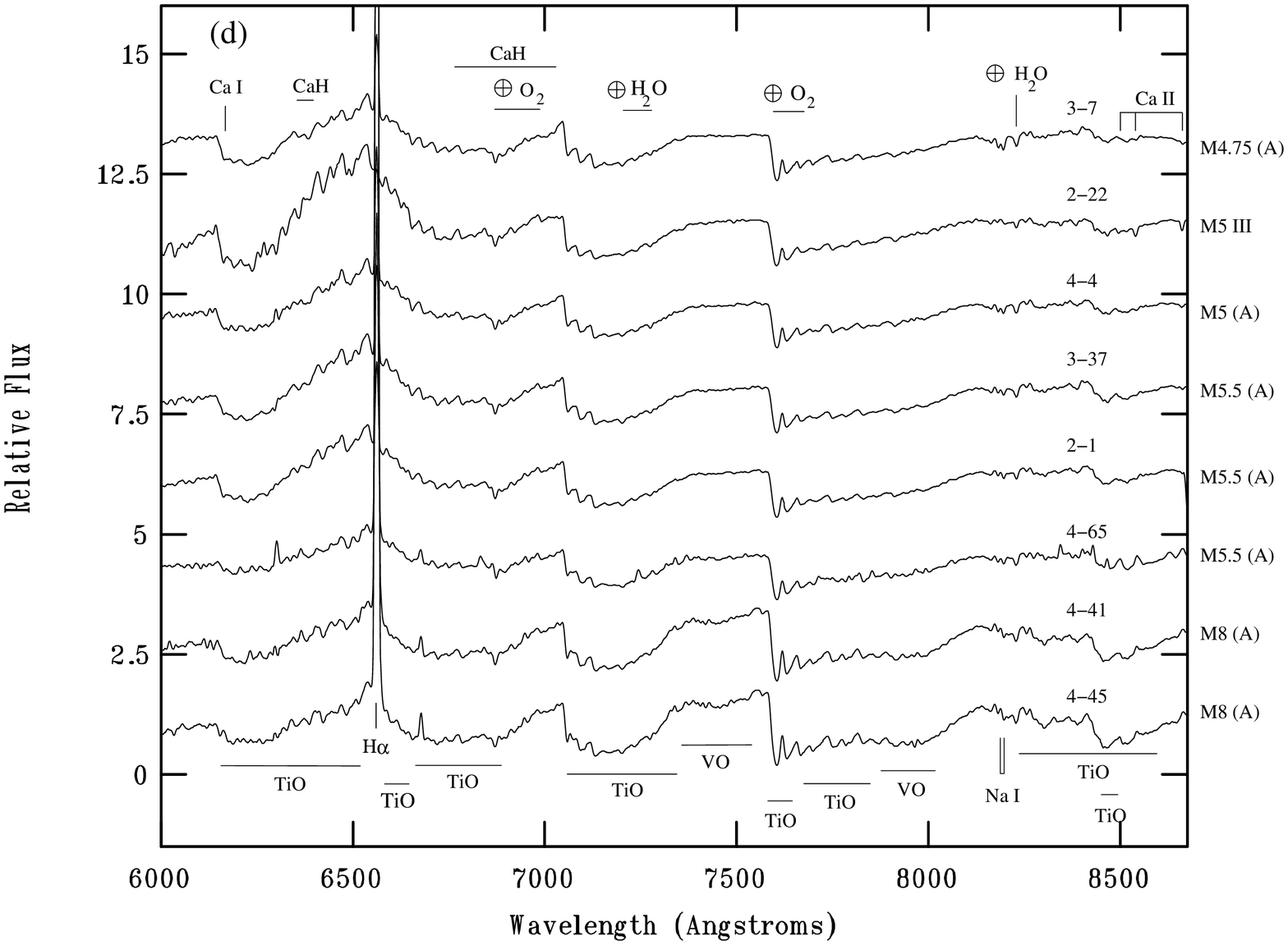}
\newpage
\centerline{FIGURE 2.A} 
\plotone{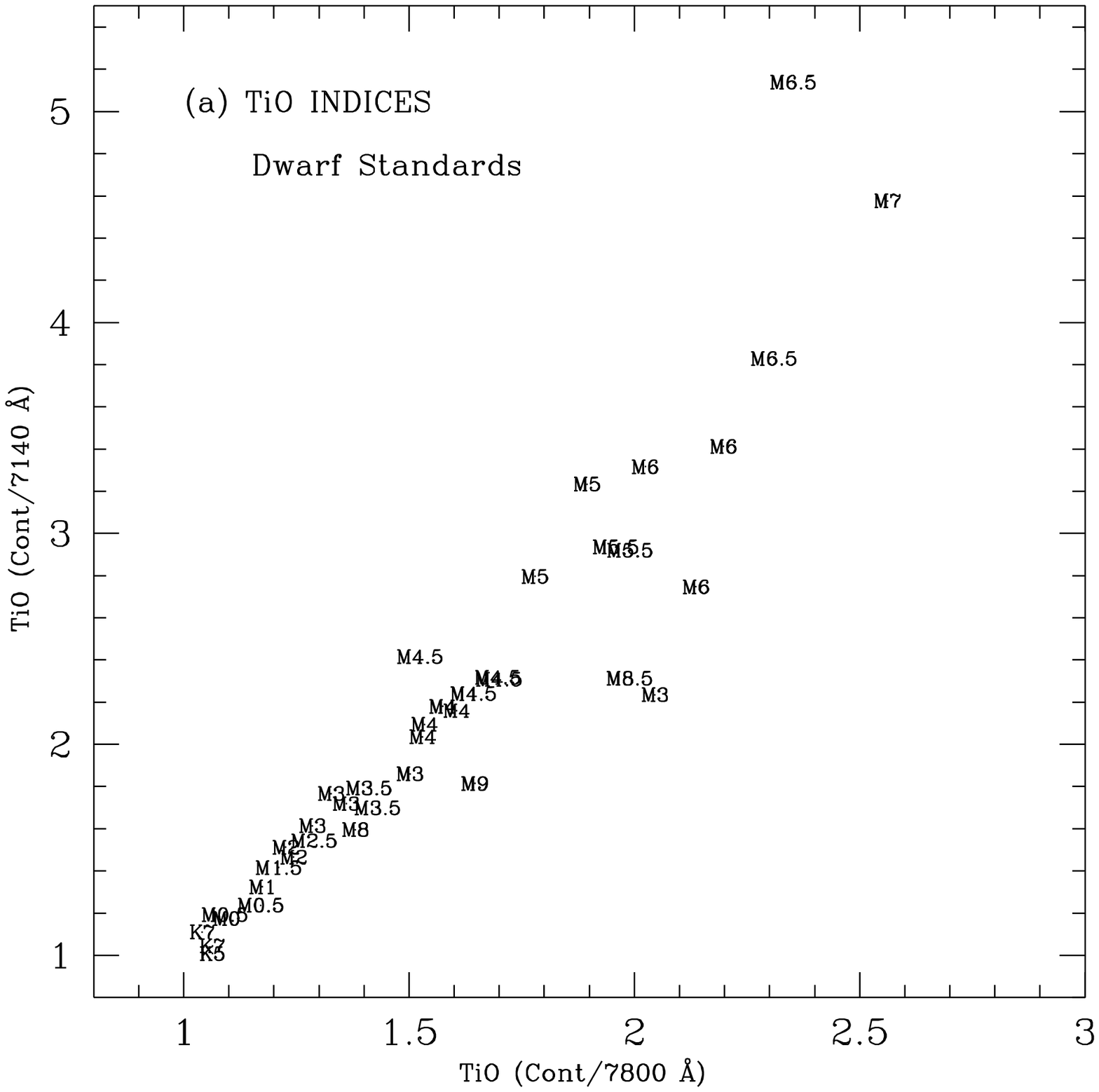}
\newpage
\centerline{FIGURE 2.B} 
\plotone{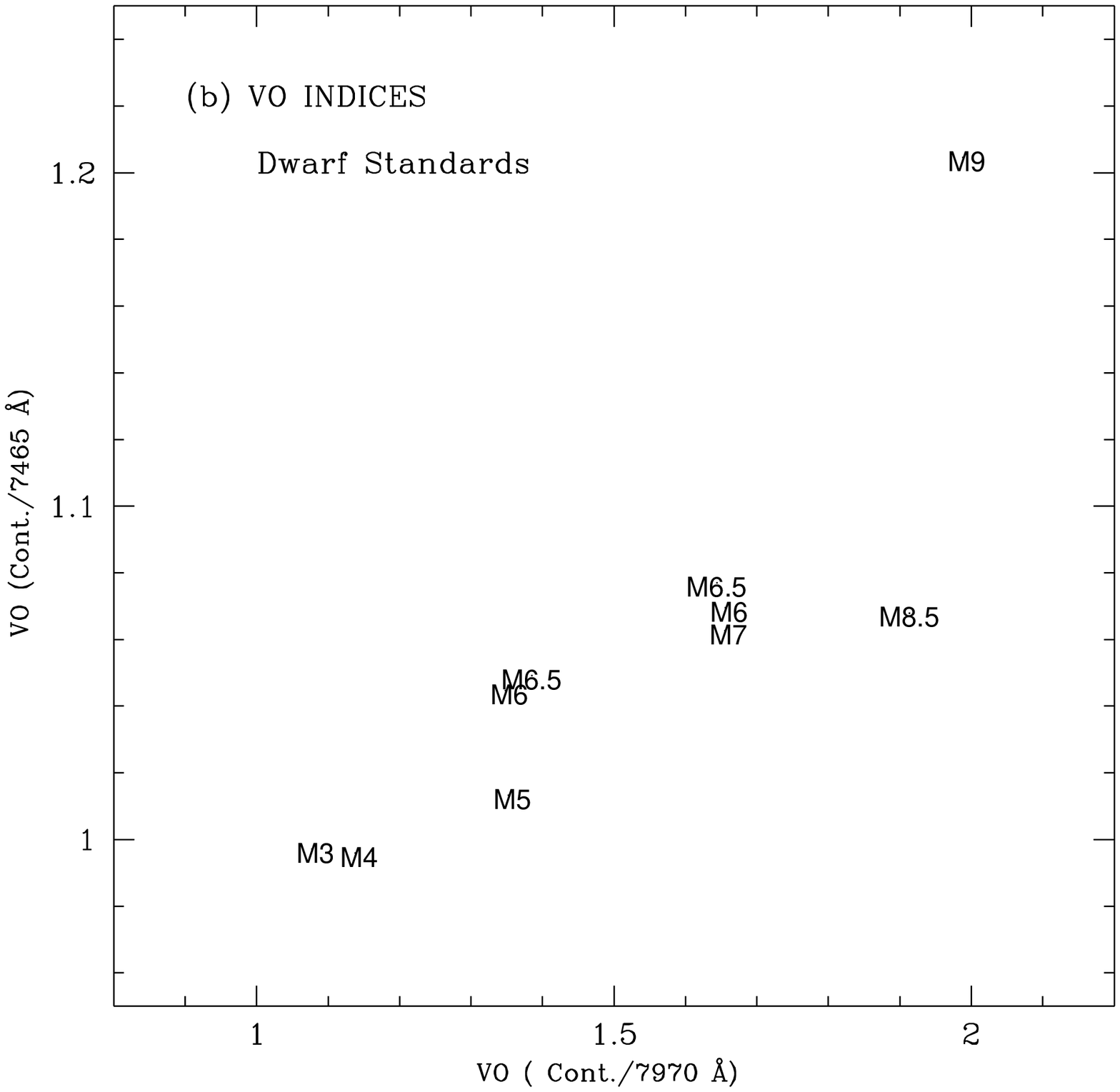}
\newpage
\centerline{FIGURE 3} 
\plotone{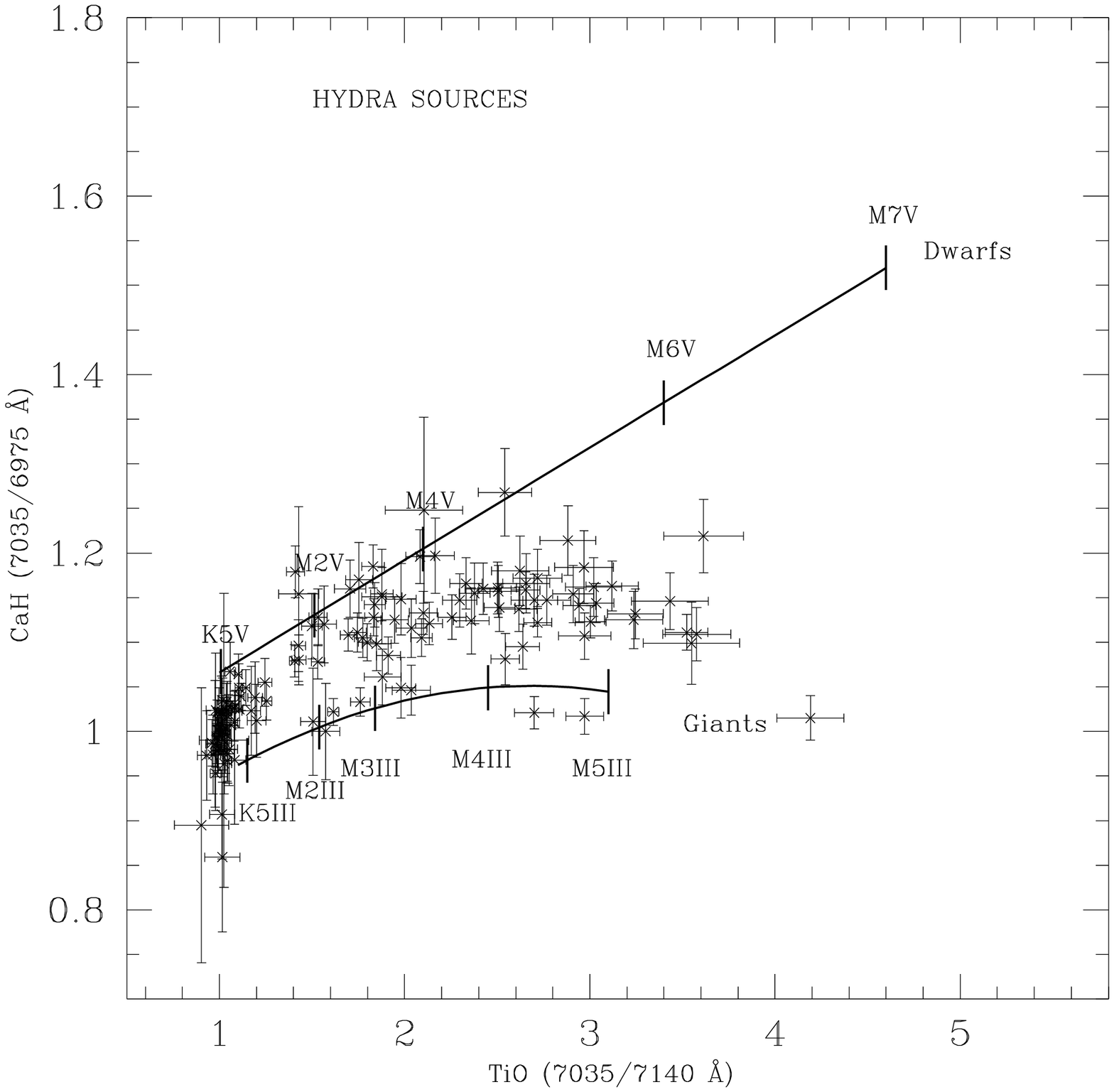}
\newpage
\centerline{FIGURE 4} 
\plotone{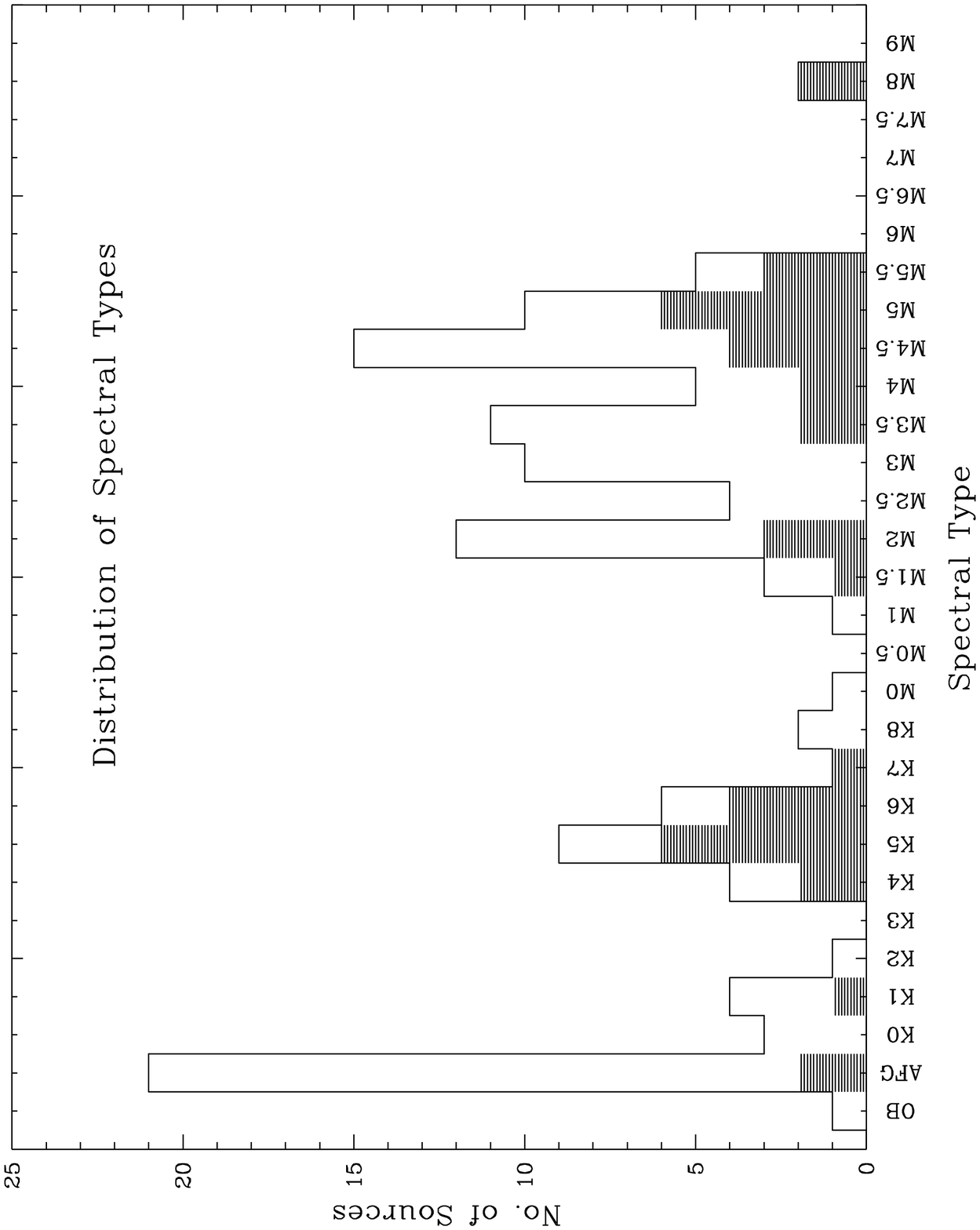}
\newpage
\centerline{FIGURE 5.A} 
\plotone{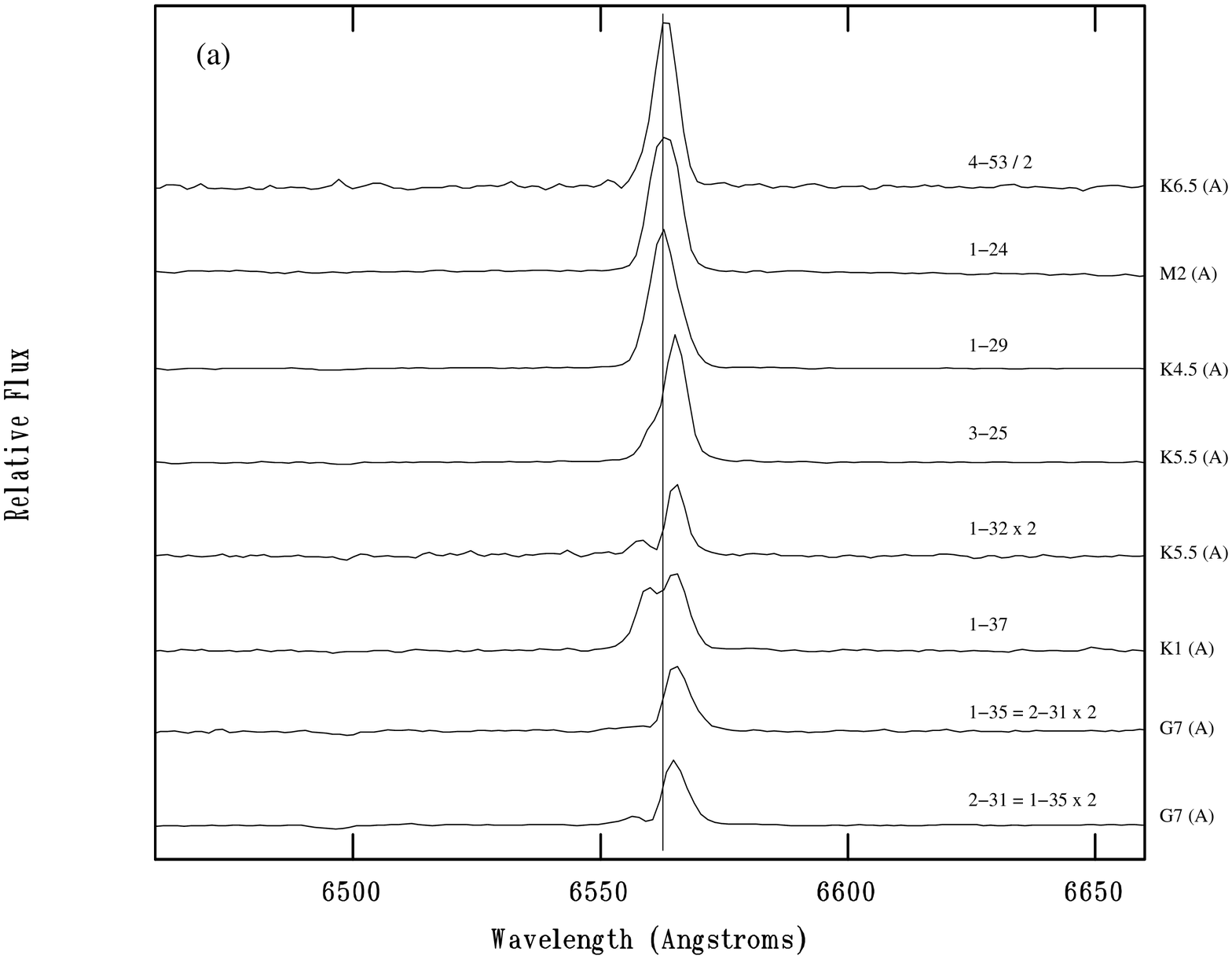}
\newpage
\centerline{FIGURE 5.B} 
\plotone{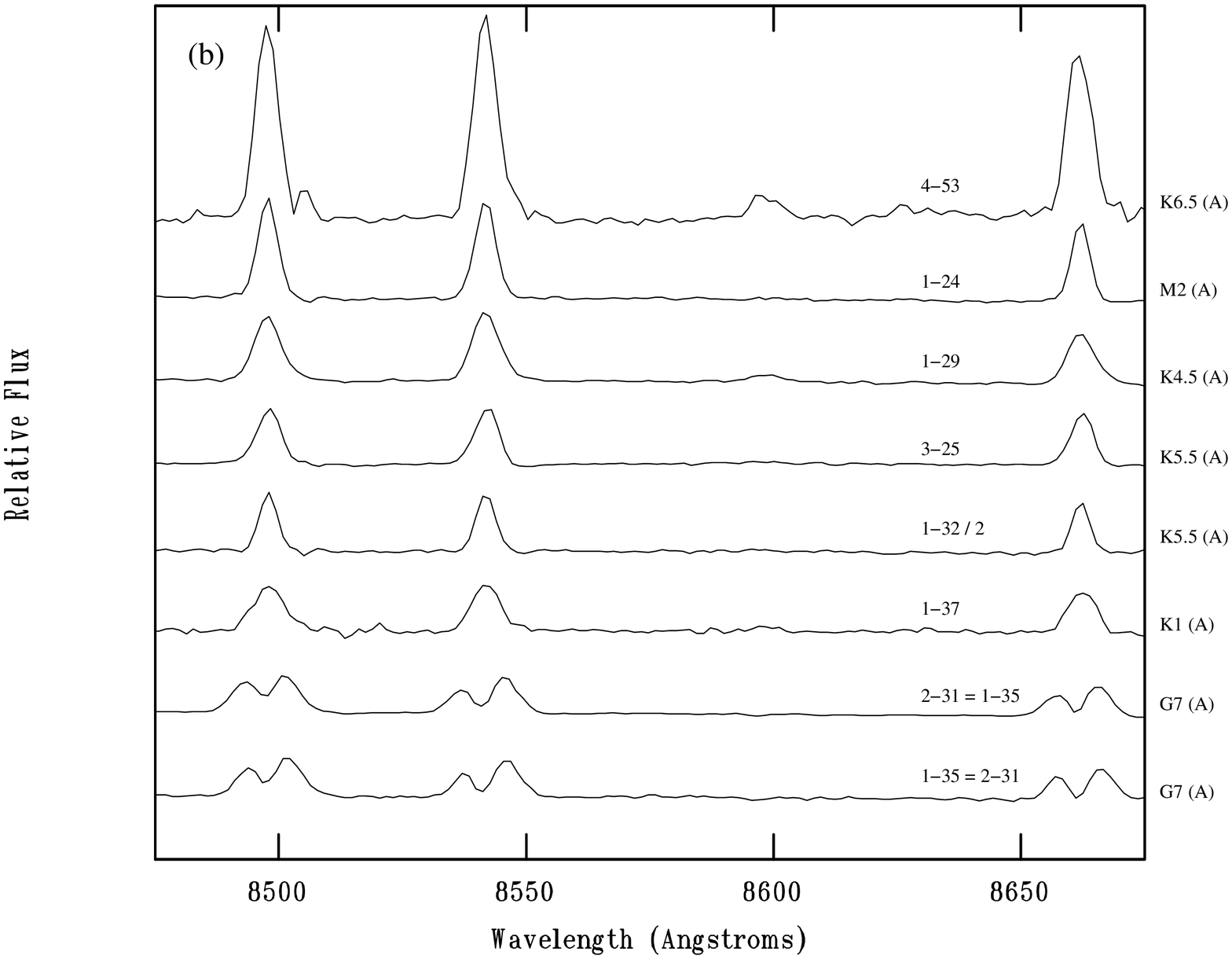}
\newpage
\centerline{FIGURE 6} 
\plotone{f6.eps}
\newpage
\centerline{FIGURE 7} 
\plotone{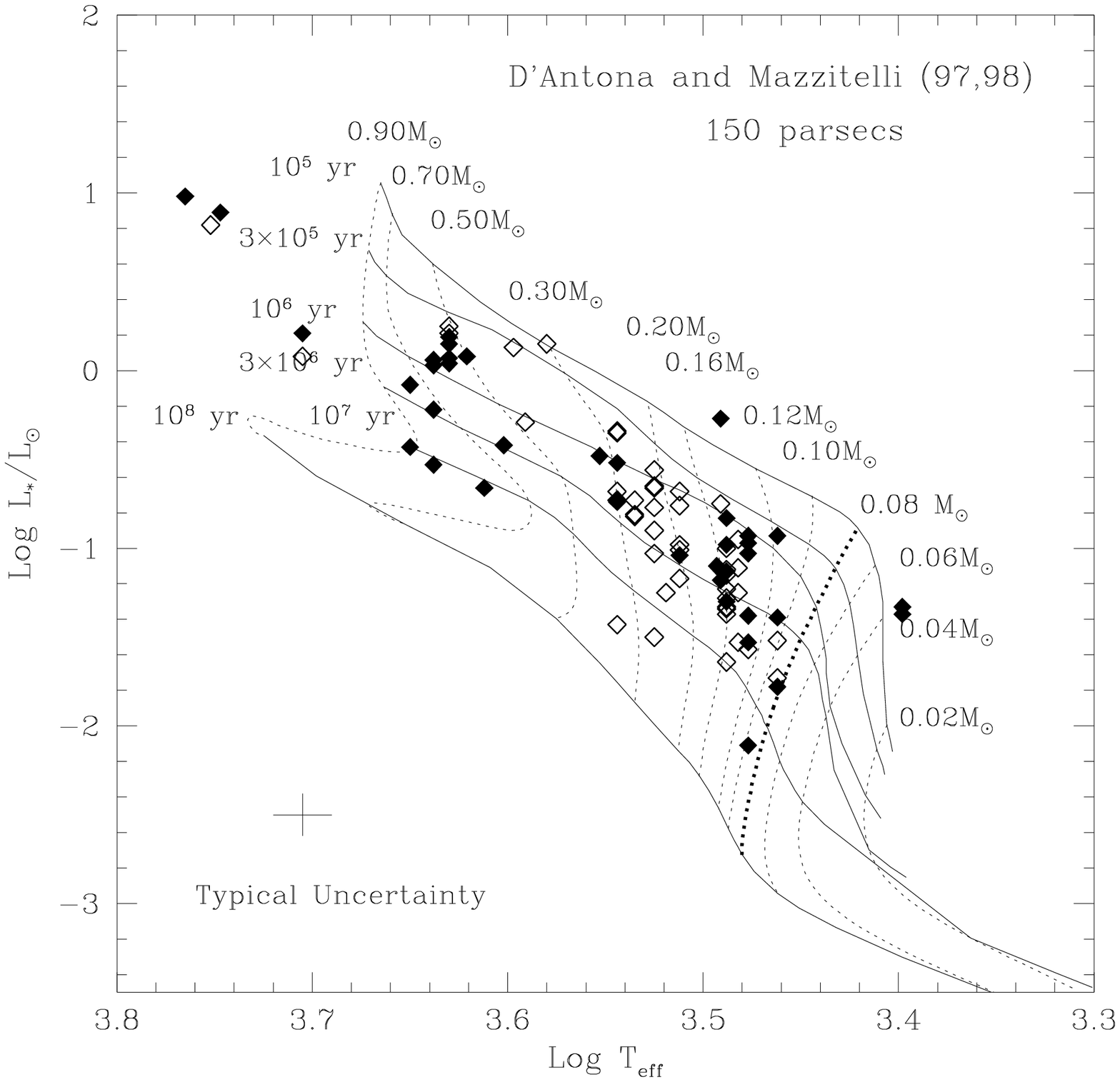}
%
%
%
%
%

\begin{references}
\reference{A96} Alcal\'a, J. M., Terranegra, L., Wichmann, R. \etal 1996, 
                Astronomy \& Astrophysics Supplement, 119, 7
\reference{A00} Alcal\'a, J. M., Covino, E., Torres, G., Sterzik, M. F., Pfeiffer, M. J., 
                \& Neuh\'auser, R. 2000, \aap, 353, 186
\reference{A96} Allen, L. E. 1996, Ph.D. Thesis, University of Massachusetts
\reference{AS95} Allen, L. E. \& Strom, K. M. 1995, \aj, 109, 1379
\reference{AM87} Andr\'e, P., Montmerle, T., \& Feigelson, E. D. 1987, \aj, 93, 1182 
\reference{AM94} Andr\'e, P. \& Montmerle, T. 1994, \apj, 420, 837
\reference{BBB02} Bate, M. R., Bonnell, I. A., \& Bromm, V. 2002, \mnras, 332, 65
\reference{B94}  Bertelli, G., Bressan, A., Chiosi, C., Fagotto, F., \& Nasi, E. 1994,
                Astronomy \& Astrophysics Supplement, 106, 275 
\reference{BCAH 98} Baraffe, I., Charbrier, G., Allard, F, \& Hauschuldt, P. H. 1998,
                   \aap, 337, 403
\reference{B91} Bessell, M. S. 1991, \aj, 101, 662
\reference{B91} Blaauw, A. 1991, in The Physics of Star Formation and Early
        Stellar Evolution, eds. C.J. Lada \& N.D. Kylafis, Kluwer Academic
        Publishers, Dordrecht,
        The Netherlands, p. 125
\reference{B01} Bontemps, S., Andr\'e, P. Kaas, A., Nordh, L., Olofsson, G., Huldtgren, M., 
                Abergel, A., Blommaert, J., Boulanger, F., Burgdorf, M., Cesarsky, C., Cesarsky, D.,
                Copet, E., Davies, J., Falgarone, E., Lagache, G., Montmerle, T., P\'erault, M.,
                Persi, P., Prusti, T., Puget, J., \& Sibille, F. 2001, \aap, 372, 173 
\reference{BA92} Bouvier, J. \& Appenzeller, I. 1992, Astronomy \& Astrophysics Supplement, 92, 481
\reference{C95} Casanova, S., Montmerle, T., Feigelson, E. D., \& Andr\'e,
                 P. 1995, \apj, 439, 752
\reference{C81} Chini, R. 1981, \aap, 99, 346
\reference{CK79} Cohen, M. \& Kuhi, L. V. 1979, \apjs, 41, 743
\reference{CFPE81} Cohen, J. G., Frogel, J. A., Persson, S. E., \& Elias,
                J. H. 1981, \apj, 249, 481
\reference{D02}  Dahn, C. et al. 2002, \aj, 124, 1170
\reference{DM97} D'Antona, F. \& Mazzitelli, I. 1997, in Cool Stars in Clusters
                 and Associations, eds. R. Pallavicini \& G. Micela,
                 Mem. S. A. It., 68, n.4.
\reference{DM98} D'Antona, F. \& Mazzitelli, I. 1998, priv. comm.
\reference{dG92} de Geus, E. J. 1992, \aap 262, 258
\reference{dZ99} de Zeeuw, P. T., Hoogerwerf, R., \& de Bruijne, J. H. J., Brown, A. G. A., 
                 \& Blaauw, A. 1999, \aj, 117, 354
\reference{DA59} Dolidze, M. V. \& Arakeylan, M. A. 1959, Soviet-Astr.--AJ, 3, 434
\reference{E87}  Edwards, S., Cabrit, S., Strom, S., Heyer, I., Strom, K., \& Anderson, E. 1987
                 \apj, 321, 473
\reference{E78} Elias, J. H. 1978, \apj, 224, 453
\reference{F96}   Feigelson, E. D. 1996, \apj, 468, 306
\reference{F98}   Festin, L. 1998, \aap, 336, 883
\reference{GSD04} Gagn\'e, M., Skinner, S. L., \& Daniel, K. J. 2004, astro-ph/0405467
\reference{GSS73} Grasdalen, G. L. Strom, K. M., \& Strom, S. E. 1973,
                 \apjl, 184, L53
\reference{GM95} Greene, T. P. \& Meyer, M. R. 1995, \apj, 450, 233
\reference{GY92} Greene, T. P. \& Young, E. T., 1992, \apj, 395, 516
\reference{G00}  Grosso, N., Montmerle, T., Bontemps, S., Andr\'e, P., \& Feigelson, E. D. 2000, \aap, 359, 113
\reference{H93}  Hartigan , P. 1993, \aj, 105, 1511
\reference{HC94} Hartmann, L., Hewett, R., \& Calvet, N. 1994, \apj, 426, 669
\reference{H02}  Hawley, S. et al. 2002, \aj, 123, 3409
\reference{H97}  Hillenbrand, L. A. 1997, \aj, 113, 1733
\reference{HW04} Hillenbrand, L. A. \& White, R. J. 2004, \apj, 616, 998
\reference{I01} Imanishi, K., Koyama, K., \& Tsuboi, Y. 2001, \apj, 557, 747
\reference{KHM91} Kirkpatrick, J. D., Henry, T. J., \& McCarthy, D. W.
                  1991, \apjs, 77, 417
\reference{LW84} Lada, C. J. \& Wilking, B. A. 1984, \aj, 287, 610
\reference{L92} Landolt, A. U. 1992, \aj, 104, 340
\reference{LFAM91} Leous, J. A., Feigelson, E. D., Andr\'e, P., \& Montmerle, T. 1991, \apj, 379, 683
\reference{L89} Loren, R. B. 1989, \apj, 338, 925
\reference{L99}  Livingston, W. C. 1999, in Astrophysical Quantities (4th Ed.),  ed. A.
                  Cox, (AIP Press: New York),  p. 341
\reference{L99} Luhman, K. L. 1999, \apj, 525, 466
\reference{LLR97} Luhman, K. L., Liebert, J., \& Rieke, G. H. 1997,
                   \apj, 489, L165
\reference{LR99} Luhman, K. L. \& Rieke, G. H. 1999, \apj, 525, 440
\reference {M97}  Mart\'{\i}n, E. L. 1997, \aap, 321, 492
\reference{MMGC98} Mart\'{\i}n, E. L., Montmerle, T., Gregorio-Hetem, J., \& Casanova, S. 1998, \mnras, 300, 733
\reference{M83} Montmerle, T., Koch-Miramonde, L., Falgarone, E., \& Grindlay, J. E. 1983, \apj, 269, 182
\reference{MCH01} Muzerolle, J., Calvet, N., \& Hartmann, L. 2001, \apj, 550, 944
\reference{N02}   Natta, A., Testi, L., Comer\'on, F., D'Antona, F., Baffa, C., Comoretto, G., \& Gennari, S.
                  2002, \aap, 393, 597
\reference{PB04}  Phelps, R. L. \& Barsony, N. 2004, \aj, 127, 420
\reference{PZ99} Preibisch, T. \& Zinnecker, H. 1999, \aj, 117, 2381
\reference{PGZ01} Preibisch, T., Guenther, E., \& Zinnecker, H. 2001, \aj, 121, 1040
\reference{P02}   Preibisch, T., Brown, A. G. A., Bridges, T., Guenther, E., \& Zinnecker, H. 2002, \aj, 124, 404 
\reference{RC01} Reipurth, B. \& Clarke, C. 2001, \aj, 122, 432
\reference{RPL96} Reipurth, B., Pedrosa, A., \& Lago, M. T. V. T. 1996, Astronomy \& Astrophysics Supplement, 120, 229
\reference{R80}  Rydgren, A. E. 1980, \aj, 85, 438
\reference{S82} Schmidt-Kaler, Th. 1982, in Landolt-Bornstein New Series,
                Numerical Data and Functional Relationships in Science and
                Technology, Group 4, Vol. 2b, edited by K. Schaffers and
                 H. H. Voigt (Springer: New York), p. 451
\reference{SD95} Sterzik, M. F. \& Durisen, R. H. 1995, \aap, 304, L9
\reference{SR49} Struve, O. \& Rudkj\"{o}bing, M. 1949, \apj, 109, 92
\reference{TD93} Torres-Dodgen, A. V. \& Weaver, W. B. 1993, \pasp, 105, 693 
\reference{V76} Vrba, F. J., Strom, S. E., Strom, K. M. 1976, \aj, 81, 958
\reference{V75} Vrba, F. J., Strom, S. E., Strom, K. M., \& Grasdalen, G. L.
                1975, \apj, 197, 77
\reference{WGM99} Wilking, B. A., Greene, T. P., \& Meyer, M. R. 1999, \aj, 117, 469
\reference{WL83} Wilking, B. A. \& Lada, C. J. 1983, \apj, 274, 698
\reference{WL89} Wilking, B. A., Lada, C. J., \& Young, E. T. 1989, \apj, 340, 823
\reference{WMG05} Wilking, B. A., Meyer, M. R., \& Greene, T. P., 2005, in preparation
\reference{WSB87} Wilking, B. A., Schwartz, R. D., \& Blackwell, J. H. 1987,
           \aj, 94, 106
\reference{WSFF98} Wilking, B., Schwartz, R. D., Fanetti, T., \&
                   Friel, E. 1997, \pasp, 109, 549
\end{references}
\end{document}